\documentclass[letterpaper, 12 pt, onecolumn, conference]{ieeeconf}

\usepackage{amsfonts}
\usepackage{amssymb,amsmath,epsfig,graphicx,amsmath,epstopdf,color}
\usepackage{mathrsfs,float}

\newtheorem{theorem}{Theorem}
\newtheorem{lemma}{Lemma}
\newtheorem{remark}{Remark}
\newtheorem{assumption}{Assumption}
\newtheorem{property}{Property}

\begin{document}

\title{Concurrent Learning Based Adaptive Control of Euler Lagrange Systems with Guaranteed Parameter Convergence}

\author{Erkan Zergeroglu, Enver Tatlicioglu and Serhat Obuz 
%\thanks{
%\newline E. Zergeroglu is with the Department of Computer Engineering, Gebze Technical University, 41400, Gebze, Kocaeli, Turkey (Email: e.zerger@gtu.edu.tr),
%\newline E. Tatlicioglu is with the Department of Electrical \& Electronics Engineering, Ege University, 35100, Bornova, Izmir, Turkey (Email. enver.tatlicioglu@ege.edu.tr).
%%}
}

%\thanks{All the correspondence should be addressed to E. Zergeroglu}

\maketitle

\begin{abstract}
This work presents a solution to the adaptive tracking control of Euler Lagrange systems with guaranteed tracking and parameter estimation error convergence. Specifically a concurrent learning based update rule fused by the filtered version of the desired system dynamics in conjunction with a desired state based regression matrix has been utilized to ensure that both the position tracking error and parameter estimation error terms converge to origin exponentially. As the regression matrix used in proposed controller makes use of the desired versions of the system states, an initial, sufficiently exciting memory stack can be formed from the knowledge of the desired system trajectory a priori, thus removing the initial excitation condition required for the previously proposed concurrent learning based controllers in the literature. The output feedback versions of the proposed method where only the position measurements are available for the controller design, (for both gradient and composite type adaptions) are also presented in order to illustrate the modularity of the proposed method. The stability and boundedness of the closed loop signals for all the proposed controllers are ensured via Lyapunov based analysis.
%Trajectory tracking control of a class of fully actuated Euler Lagrange systems is considered in this work. System dynamics is considered to be subject to parametric uncertainties and on--line identification uncertain model parameters is also aimed. When compared with the relevant past research, via a novel approach, desired states are proposed to be used in forming the regression matrix and a desired compensation based concurrent learning type adaptive update rule is designed. Via utilizing novel Lyapunov analysis, semi--global exponential convergence of both tracking and parameter identification error to the origin is ensured. 
%
%In the second part, the control design is further restricted by the unavailability of velocity sensing. To remedy this, a surrogate signal is designed to cover for the lack of velocity measurements. Two different adaptive update laws, one gradient based and one least squares type with time--varying gain matrix, both of which being fused with concurrent learning type rules are proposed. Via novel Lyapunov tools, convergence of the tracking error to the origin while simultaneously identifying uncertain model parameters are ensured. 
\end{abstract}

\section{Introduction}

When the dynamical model of the system to be controlled is subject to parametric/structured uncertainties, adaptive methods are in order. A long--standing research question in adaptive control is whether accurate identification of uncertain model parameters is achieved when meeting the tracking or regulation objective \cite{Sastry-Boston}, \cite{sun}, \cite{KrsticBook}. Provided the regression matrix satisfies a persistent excitation (PE) condition, uncertain parameters may be identified \cite{Sastry-Boston}, \cite{sun}. Mathematically speaking, the PE condition requires the regression matrix to variate in a time window of $[t,t+T]$ for all $t$ for some positive $T$. Both guaranteeing PE condition and checking whether the PE condition is satisfied or not are tough tasks. While there are some past works to relax the need for PE condition, an interesting idea was presented in \cite{Chowdhary2010}, \cite{Chowdhary2010a}. While removing the need for PE condition via forming a matrix by utilizing data starting from initial time to the current time, the proposed method required time derivative of the state vector to be available. To remedy this, in \cite{Kamalapurkar17}, a state derivative was made use of and later in \cite{Kamalapurkar18}, a modification is proposed that included an integrator in the adaptive update rule which resulted in not needing time derivative of the state vector. Though the need of state derivative has been  removed by this method, the problem of controlling the overall systems by only using the output (\textit{i.e.} output feedback control (OFB)) remained unresolved.
%While state derivative is not required anymore, when control of Euler Lagrange (EL) systems is considered any realistic controller should depend only on the system output (\textit{i.e.}, position measurements) and not even on velocity measurements. 

In this work, the concurrent learning based adaptive control method is aimed to be improved to require only output measurements when controlling fully actuated EL systems. To reach this aim, first a desired compensation adaptive law (DCAL) version of the concurrent learning based adaptive control is proposed when full state feedback (\textit{i.e.}, both position and velocity measurements) is available. After achieving similar results to that of concurrent learning based adaptive controllers in the literature, the proposed controller is extended to the output feedback case for two different adaptive update rules. Via rigorous stability analysis, all three controllers (one FSFB and two OFB) are proven to ensure tracking objectives while simultaneously identifying uncertain dynamical system parameters.

\section{Dynamic Model and Properties}\label{model}

The mathematical model of an $n$ degree of freedom direct drive robot manipulator is considered to have the following form \cite{lewis}, \cite{spong-book}
\begin{equation}
M\left(q\right)\ddot{q} + C\left(q ,\dot{q}\right)\dot{q} + G\left(q \right) + F_{d}\dot{q} = \tau \label{modal}
\end{equation}
where $q \left(t\right)$, $\dot{q}\left(t\right)$, $\ddot{q}\left(t\right)\in \mathbb{R}^{n}$ denote joint position, velocity, and acceleration vectors, respectively, $M(q)\in \mathbb{R}^{n\times n}$ represents positive--definite and symmetric inertia matrix, $C\left(q ,\dot{q}\right) \in \mathbb{R}^{n\times n}$ denotes centripetal--Coriolis terms, $G(q)\in \mathbb{R}^{n}$ is the vector representing the gravitational effects, $F_{d}\in \mathbb{R}^{n\times n}$ is the constant, positive--definite, diagonal viscous friction matrix, and $\tau\left(t\right)\in \mathbb{R}^{n}$ denotes the control input torque. The dynamical terms in \eqref{modal} satisfy the following standard properties that will be utilized in the subsequent analysis.

\begin{property} \label{prop1}
The inertia matrix $M\left(q\right)$ satisfies following inequalities \cite{spong-book}
\begin{equation}
m_{1} \Vert \xi \Vert^2 \leq \xi^T M\left(q\right) \xi \leq m_{2} \Vert \xi \Vert^2 \quad \forall \quad  \xi \in \mathbb{R}^n \label{prp_1}
\end{equation} 
where $m_{1}$, $m_{2}\in\mathbb{R}$ are known, positive bounding constants, and $\Vert \cdot \Vert$ denotes the standard Euclidean norm of a vector.
\end{property}

\begin{property} \label{prop2}
The time derivative of the inertia matrix, $ \dot{M}(q)$, can be written in terms of the centripetal--Coriolis matrix as
\begin{equation}
\dot{M}(q) = C(q, \dot{q}) + C^T (q, \dot{q}) \label{prp_2}
\end{equation}
and they satisfy the following relationship \cite{spong-book}
\begin{equation}
\xi^{T} \left(\dot{M}-2 C\right) \xi = 0 \quad \forall \quad \xi \in \mathbb{R}^{n} . \label{prp_2_1}
\end{equation}
\end{property}

\begin{property} \label{prop3}
The centripetal--Coriolis matrix satisfies the following switching property \cite{lewis}
\begin{equation}
C\left(q,\nu\right)\xi =C\left(q,\xi\right)\nu \quad \forall \quad \nu \text{, } \xi \in\mathbb{R}^{n} .\label{ekle10}
\end{equation}
\end{property}

\begin{property} \label{DynamicBounds}
The dynamical terms in \eqref{modal} satisfy the following bounds \cite{lewis} 
\begin{eqnarray}
\left\Vert M(\xi)-M(\nu)\right\Vert_{i\infty} &\leq & \zeta_{M1}\left\Vert \xi-\nu \right\Vert \label{bound1a} \\
%\left\Vert M^{-1}(\xi)-M^{-1}(\nu)\right\Vert_{i\infty} &\leq & \zeta_{M2}\left\Vert \xi-\nu \right\Vert \label{notation3} \\
\left\Vert C\left(q,\xi\right)\right\Vert_{i\infty} &\leq &\zeta_{C1}\left\Vert \xi \right\Vert \\ \label{CCC_eq1}
\left\Vert C(\xi,w)-C(\nu,w)\right\Vert_{i\infty} &\leq & \zeta_{C2} \left\Vert \xi-\nu \right\Vert \left\Vert w\right\Vert \label{bound1b} \\
\left\Vert G(\xi)-G(\nu)\right\Vert &\leq & \zeta_{G}\left\Vert \xi-\nu \right\Vert  \label{ekle13} \\
\left\Vert F_{d}\right\Vert_{i\infty}&\leq & \zeta_{F}\label{CCC_eq2}
\end{eqnarray}
$\forall$ $\xi$, $\nu$, $w\in\mathbb{R}^{n}$, where $\zeta_{M1}$, $\zeta_{M2}$, $\zeta_{C1}$, $\zeta_{C2}$, $\zeta_{G}$, $\zeta_{F} \in\mathbb{R}$ are known, positive bounding constants and $\left\Vert \cdot \right\Vert_{i\infty}$ denotes induced infinity norm of a matrix. 
\end{property}

\begin{property} \label{P5}
The left hand side of \eqref{modal} can be reconfigured to be written in linearly parameterized form as
\begin{equation}
Y\left(q,\dot{q},\ddot{q}\right)\theta = M\left( q\right)\ddot{q} + C\left( q,\dot{q}\right)\dot{q} + G\left( q\right) + F_{d}\dot{q} \label{prop5}
\end{equation}
with $Y\left(q,\dot{q},\ddot{q}\right)\in \mathbb{R}^{n\times p}$ being a regression matrix and $\theta \in \mathbb{R}^{p}$ containing constant model parameters that depends on physical properties of the robot manipulator. Desired form of \eqref{prop5} can be written as
\begin{equation}
Y_d\left(q_{d},\dot{q}_{d},\ddot{q}_{d}\right)\theta = M\left( q_{d}\right)\ddot{q}_{d} + C\left( q_{d},\dot{q}_{d}\right)\dot{q}_{d} + G\left( q_{d}\right) + F_{d}\dot{q}_{d} \label{prop5d}
\end{equation}
in which $Y_d\left(q_{d},\dot{q}_{d},\ddot{q}_{d}\right)\in \mathbb{R}^{n\times p}$ is the desired version of the regression matrix that is a function of the desired joint position, velocity and acceleration vectors, denoted respectively with $q_{d}\left( t\right)$, $\dot{q}_{d}\left( t\right)$, $\ddot{q}_{d}\left( t\right) \in \mathbb{R}^{n}$.
\end{property}

\begin{assumption} \label{Ass1}
The subsequent design and analysis relies on the assumption that the robot manipulator is initially at rest (\textit{i.e.}, $\dot{q} \left(0\right) = 0_{n\times 1}$) and the desired trajectory is chosen such that $\dot{q}_{d} \left(0\right) = 0_{n\times 1}$.
\end{assumption}

\section{Full--state Feedback Control}\label{fsfb}

The main control objective is to ensure that $q\left( t\right)$ tracks $q_{d}\left( t\right)$ under the restriction that the dynamic model in \eqref{modal} is subject to parametric uncertainties; that is, $\theta$ introduced in Property \ref{P5} is uncertain. Identification of the uncertain dynamic model parameters is also aimed. Closed loop stability must be ensured as well. The controller scheme developed in this section requires both joint position and velocity measurements for implementation.

\subsection{Error System Development \& Controller Design}\label{control}

To quantify the tracking control objective, the joint position tracking error signal, denoted by $e\left(t\right)\in \mathbb{R}^{n}$, is defined as the difference between desired and actual joint positions as
\begin{equation}
e \triangleq q_{d}-q  \label{e}
\end{equation}
and the filtered tracking error term, $r(t) \in \mathbb{R}^n$, is defined as
\begin{equation}
r \triangleq \dot{e}+\alpha e \label{r} 
\end{equation}
where $\alpha \in \mathbb{R}$ is a positive gain. Also, in order to quantify the performance of the adaptation algorithm, parameter estimation error, shown with $\tilde{\theta}(t) \in \mathbb{R}^p$, is defined as the difference between the actual and estimated system parameters in the form
\begin{equation}
\tilde{\theta} = \theta - \hat{\theta} \label{thetaTilde}
\end{equation}
where $\hat{\theta}(t) \in \mathbb{R}^p$ denotes the yet to be designed adaptively updated estimation of uncertain system parameter vector $\theta$.

In order to facilitate the construction of a desired state based estimation for the parameter update design,  we apply  a filtering technique on the control input torque in the following form
\begin{equation}
\tau_f = f * \tau \label{tf}
\end{equation}
where  $\tau_f(t) \in \mathbb{R}^n $ is the filtered version of the control input torque $\tau(t)$, ``$*$'' denotes the standard convolution operation and the filter function $f(t) \in \mathbb{R}$ is designed as
\begin{equation}
f = \beta \exp(-\beta t) \label{f_filter} 
\end{equation}
with $\beta \in \mathbb{R}$ being a constant, positive filter parameter. In \eqref{tf}, inserting \eqref{modal} for $\tau(t)$ and utilizing properties of the convolution operation, we can construct the filtered version of the input signal as
\begin{equation}
\tau_f = Y_f \theta \label{tf_param}
\end{equation}
where $Y_f (q, \dot{q}) \in \mathbb{R}^{n \times p}$ is the filtered version of the regression matrix introduced in \eqref{prop5} which does not require acceleration measurements and is explicitly given as
\begin{eqnarray}
Y_f \theta &=& \dot{f}(t) * \left[ M \left( q(t)\right) \dot{q}(t) \right] \nonumber \\ 
&& + f(0) M \left( q(t) \right) \dot{q}(t) - f(t) M \left( q(0) \right) \dot{q}(0) \nonumber \\
&& + f(t) * \left[ - \dot{M}\left( q(t) \right) \dot{q}(t) + C\left( q(t), \dot{q}(t)\right) \dot{q}(t) + G(q(t)) + F_d \dot{q}(t) \right] \label{YfQ}
\end{eqnarray}
%/************************ \\
%Ben üstteki denklemde hatalar olduğuna inanıyorum. Ama şimdilik detaylı bakamadım. Arxiv'de böyle kalır sonra düzeltiriz... \\
%******************************************************** \\
where \eqref{prp_2} of Property \ref{prop2} was used and the time derivative of the filter signal in \eqref{f_filter} is given as
\begin{equation}
\dot{f} = - \beta^2 \exp(-\beta t) . \label{f_dot}
\end{equation}
At this stage, similar to \cite{compositeRob99}, in order to remove the state measurement dependency in the parameter update rule design, we reformulate \eqref{YfQ} to obtain its desired state dependent version as
\begin{eqnarray}
Y_{df} \theta &=& \dot{f}(t) * \left[ M\left( q_d(t)\right) \dot{q}_d(t) \right] \nonumber \\ 
&& + f(0) M\left( q_d(t)\right) \dot{q}_d(t) - f(t) M\left( q_d(0)\right) \dot{q}_d(0) \nonumber \\ 
&& + f(t) * \left[ - \dot{M}\left( q_d(t) \right) \dot{q}_d(t) + C\left( q_d(t), \dot{q}_d(t)\right) \dot{q}_d(t) + G(q_d(t)) + F_d \dot{q}_d(t) \right] \label{YdfQ}
\end{eqnarray}
where $Y_{df}(q_d, \dot{q}_d) \in \mathbb{R}^{n \times p}$ is the desired filtered regression matrix. We are now ready to define a prediction error type signal, represented with $\varepsilon (t) \in \mathbb{R}^n$, as 
\begin{equation}
\varepsilon = \tau_f - Y_{df} \hat{\theta} \label{epsi01}
\end{equation}
where $Y_{df}$ and $\hat{\theta}$ were introduced in \eqref{YdfQ} and \eqref{thetaTilde}, respectively. It is noted that, \eqref{epsi01} can be reformulated to reach
\begin{eqnarray}
\varepsilon &=& Y_f \theta - Y_{df} \theta + Y_{df} \theta - Y_{df} \hat{\theta} \nonumber \\ 
&=& \Omega + Y_{df} \tilde{\theta} \label{epsi02}
\end{eqnarray}
with the auxiliary error--like term $\Omega(t) \in \mathbb{R}^n$ being defined as
\begin{equation}
\Omega \triangleq Y_f \theta - Y_{df} \theta \label{Omega}
\end{equation}
which, in view of \eqref{YfQ} and \eqref{YdfQ}, can be reformulated as
\begin{eqnarray}
\Omega &=& \dot{f} * \left[ M(q) \dot{q} - M(q_d) \dot{q}_d \right] \nonumber \\ 
&& + f(0) \left[ M (q) \dot{q} - M(q_d) \dot{q}_d \right] \nonumber \\ 
&& - f(t) \left[ M(q(0)) \dot{q}(0) - M(q_d(0))\dot{q}_d(0) \right] \nonumber  \\ 
&& + f(t) * \left[ - \dot{M}\left( q \right) \dot{q} + \dot{M}\left( q_d \right) \dot{q}_d + C(q, \dot{q}) \dot{q} - C(q_d, \dot{q}_d) \dot{q}_d + G(q) - G(q_d) + F_d \left( q- q_d \right) \right] \label{Omega2}
\end{eqnarray}
where the dependency of some of the mathematical quantities on time were dropped whenever it is trivial. The auxiliary error--like term $\Omega$ can be upper bounded as (details are given in Appendix \ref{appremark})
\begin{equation}
\Vert \Omega \Vert \leq \rho_1 (\Vert x \Vert) \Vert r \Vert \label{OmegaBoundFsfb}
\end{equation}
where $\rho_1 (\cdot) \in \mathbb{R}$ is a known, positive bounding function and $x \left(t\right) \triangleq \left[ e^T \quad r^T \right]^T \in \mathbb{R}^{2n}$. 

\begin{remark}
From the definition of the filtered version of the control input torque in \eqref{tf} and the filter function in \eqref{f_filter}, the relation between the control input torque $\tau$ and its filtered version $\tau_f$ can, with an abuse of notation, be written in ``$s$'' domain as
\begin{equation}
s \tau_f + \beta \tau_f = \beta \tau  
\end{equation} 
which in time domain would correspond to the following differential relationship
\begin{equation}
\dot{\tau}_f + \beta \tau_f = \beta \tau \quad \tau_f(0) = 0_{n\times 1} . \label{tauf_dif}
\end{equation}
Therefore, the filtered version of the control input torque vector, $\tau_f\left(t\right)$, can be calculated via
\begin{equation}
\tau_f (t) = \beta \exp\left(-\beta t \right) \int_0^t \exp\left(\beta \sigma\right) \tau(\sigma) d\sigma \label{ahaInt01}
\end{equation} 
and similarly, the filtered version of the desired regression matrix, $Y_{df} (t)$, can be evaluated using
\begin{equation}
Y_{df}(t) = \beta \exp\left(-\beta t \right) \int_0^t \exp\left( \beta \sigma\right) Y_d(\sigma) d\sigma . \label{ahaInt02}
\end{equation}
\end{remark}

%\subsection{Controller Formulation}\label{fsfb}

Taking the time derivative of \eqref{r}, pre--multiplying the resultant by $M(q)$, substituting for \eqref{modal} and then adding / subtracting the $C(q,\dot{q})r$ and $Y_d \theta$ terms, we reach
\begin{equation}
M(q) r = -C(q,\dot{q}) r + Y_d \theta + \tilde{W} - \tau \label{OpenloopR}
\end{equation} 
where the auxiliary term $\tilde{W} \left(q,\dot{q},q_d,\dot{q}_d,\ddot{q}_d \right) \in \mathbb{R}^n$ is defined as
\begin{equation}
\tilde{W} \triangleq M(q) \left( \ddot{q}_d +\alpha \dot{e}\right) + C(q, \dot{q}) \left( \dot{q}_d +\alpha e\right) + G(q) + F_d \dot{q} - Y_d \theta . \label{Wtilde}
\end{equation}
Note that, the norm of the expression in \eqref{Wtilde} can be upper bounded as follows \cite{marcio-book}
\begin{equation}
\Vert \tilde{W} \Vert \leq \rho_2 (\Vert x \Vert) \Vert x \Vert  \label{WtildeBound}
\end{equation} 
where $\rho_2 \left( \cdot \right) \in \mathbb{R}$ is a known, positive bounding function. 

Based on the open loop error dynamics and the subsequent stability analysis, we propose the following control input torque
\begin{equation}
\tau = Y_d \hat{\theta} + K_p e + K_r r + v_r\label{controlInput}
\end{equation}
where $K_e$, $K_r \in \mathbb{R}^{n \times n}$ are positive definite, diagonal control gain matrices, $v_r(t) \in \mathbb{R}^n$ is an auxiliary controller yet to be designed, and the parameter estimation vector $\hat{\theta}(t)$ is computed via the following update rule 
\begin{equation}
%\dot{\hat{\theta}} = \Gamma Y_d^T r + k_{\theta} \Gamma \sum_{i=1}^{N} (Y_{df})_i^T \biggl( (t_f)_i - (Y_{df})_i \hat{\theta}  \biggr) 
\dot{\hat{\theta}} = \Gamma Y_d^T r + k_{\theta} \Gamma \sum_{i=1}^{N} \mathcal{Y}_i^T \biggl( \mathcal{U}_i - \mathcal{Y}_i \hat{\theta}  \biggr) \label{thetaCap}
\end{equation}
where $\Gamma \in \mathbb{R}^{p \times p}$ is a diagonal, positive definite, constant adaptation gain matrix, $k_{\theta} \in \mathbb{R}$ is a positive adaptation gain and $\mathcal{Y}_i \triangleq \mathcal{Y}(t_i)$, $\mathcal{U}_i \triangleq \mathcal{U}(t_i)$, $\mathcal{Y} : \left[0, \infty \right)\rightarrow \mathbb{R}^{n \times p} $  and $\mathcal{U}:\left[0, \infty \right)\rightarrow \mathbb{R}^{n}$ are defined as
\begin{eqnarray}
\mathcal{Y}(t_i) & \triangleq & Y_{df} (t_i) = \beta \exp\left(-\beta t_i \right) \int_0^{t_i} \exp\left(\beta \sigma\right) Y_{d}(\sigma) d\sigma \label{aha01} \\
\mathcal{U}(t_i)  & \triangleq & \tau_f (t_i)= \beta \exp\left(-\beta t_i \right) \int_0^{t_i} \exp\left(\beta \sigma\right) \tau(\sigma) d\sigma \label{aha02} .
\end{eqnarray}
Using the definitions given in \eqref{tf_param}, \eqref{epsi02}, \eqref{aha01} and \eqref{aha02}, the summation in \eqref{thetaCap} can be reformulated to have the following form
\begin{equation}
\dot{\hat{\theta}} = \Gamma Y_d^T r + k_{\theta} \Gamma \sum_{i=1}^{N} \mathcal{Y}_i^T \biggl( \Omega_i + \mathcal{Y}_i \tilde{\theta} \biggr) . \label{aha03}
\end{equation}
 
The closed loop dynamics for $r(t)$ can be obtained by substituting the control input torque in \eqref{controlInput} into \eqref{OpenloopR} as
\begin{equation}
M(q) \dot{r} = -C(q,\dot{q})r +Y_d \tilde{\theta} +\tilde{W} - K_p e - K_r r - v_r \label{clR} 
\end{equation} 
and the dynamics for the parameter estimation error is obtained via substituting \eqref{aha03} into the time derivative of \eqref{thetaTilde} to reach
\begin{eqnarray}
\dot{\tilde{\theta}} &=& - \dot{\hat{\theta}} \nonumber \\ 
&=& - \Gamma Y_d^T r - k_{\theta} \Gamma \sum_{i=1}^{N} \mathcal{Y}_i^T \Omega_i - k_{\theta} \Gamma \sum_{i=1}^{N} \mathcal{Y}_i^T \mathcal{Y}_i \tilde{\theta} . \label{thetaCl}
\end{eqnarray}
We would like to note that the summation in the second expression of \eqref{thetaCl} can be upper bounded as
\begin{eqnarray}
\left\Vert \sum_{i=1}^{N} \mathcal{Y}_i^T \Omega_i \right\Vert = \left\Vert \sum_{i=1}^{N} Y_{df}^{T} \left(t_i\right) \Omega \left(t_i\right) \right\Vert &\leq & \sum_{i=1}^{N} \Vert Y_{df} \left(t_i\right) \Vert \Vert \Omega \left(t_i\right) \Vert \nonumber \\ 
&\leq & \sum_{i=1}^{N} \Vert Y_{df} \left(t_i\right) \Vert \rho_1 \left( \Vert x\left(t_i\right) \Vert \right) \Vert r \left(t_i\right) \Vert \nonumber \\ 
&\leq & \lambda_d N \sup_{i=1,\cdots,N} \lbrace \rho_1 \left( \Vert x\left(t_i\right) \Vert \right) \rbrace \Vert r \Vert \nonumber \\
&\leq & \lambda_d N \overline{\rho}_1 \Vert r \Vert \label{boundYdfS}
\end{eqnarray} 
where $\lambda_{d} \in \mathbb{R}$ is a positive bounding constant satisfying
\begin{equation}
\Vert Y_{df} \left(t_i\right) \Vert \leq \lambda_d \, \forall t_i ,
\end{equation}
and the bounding function $\overline{\rho}_1 \in \mathbb{R}$ is defined as follows 
\begin{equation}
\overline{\rho}_1 \triangleq \sup_{i=1,\cdots,N} \lbrace \rho_1 \left( \Vert x\left(t_i\right) \Vert \right) \rbrace .
\end{equation}
It is noted that, for the summation in third expression of \eqref{thetaCl}, there exists $\overline{\lambda}$, $\underline{\lambda} \in \mathbb{R}$ such that
\begin{eqnarray}
\overline{\lambda} &\geq& \lambda_{\max} \left\lbrace \sum_{i=1}^{N} \mathcal{Y}_i^T \mathcal{Y}_i \right\rbrace , \label{overLamda} \\
\underline{\lambda} &\leq& \lambda_{\min} \left\lbrace \sum_{i=1}^{N} \mathcal{Y}_i^T \mathcal{Y}_i \right\rbrace \label{underLamda}
\end{eqnarray}
where $\lambda_{\min} \lbrace \cdot \rbrace$, $\lambda_{\max} \lbrace \cdot \rbrace$ are used to represent the minimum and maximum eigenvalues of a matrix.

%/************************ \\
%Burada history stack hakkinda bilgi vermem ve hitli adamı cite etmem gerekiyor... belki extra bir remark daha eklenebilinir... history stack desired statelerden oluştuğu için off-line da başlatılabilir bu sebeple IE ye de ihtiyaç duymadan bu işi halledebilir denmesi gerekiyor.... \\
%******************************************************** \\

%\begin{remark}
%It is known that improved performance is obtained with identification algorithms that use sensor measurements simply because of the presence of noise in these measurements. On the other hand, for applications that the desired trajectory is available \textit{a priori}, utilizing desired states instead of state measurements allows the data matrix $\mathcal{Y}$ to be formed off--line and thus the condition of initial excitation may be shown to exist in advance.
%\end{remark}

\begin{remark}
As stated in Assumption 1 of \cite{Kamalapurkar18}, the controller formulation of \cite{Kamalapurkar18} requires the system states be sufficiently excited over a finite period time ( until $\sum_{i=1}^N \mathcal{Y}_i^T \mathcal{Y}_i $ is full rank).  In order to achieve initial excitation, the authors of \cite{Kamalapurkar18} proposed to add  extra perturbation to the desired system states, which in turn would indirectly excite the actual system states, for a finite time, $T$, before ensuring that sufficient data has been collected to learn system parameters and presented a two stage stability analysis (one stage for  $0 \leq t \leq T$ and second stage for $t\geq T$ where $t$ is the time).

Since the controller formulation in this work make use of desired system states, the entries $\mathcal{Y}_i$ given in \eqref{aha01} can be calculated and selected off-line to ensure $\lambda_{min}\lbrace \sum_{i=1}^N \mathcal{Y}_i^T \mathcal{Y}_i \rbrace \geq \underline{\lambda}$ for some positive valued $\underline{\lambda}$ (this argument is still be satisfied  even when the desired trajectory of the system is partially known) removing the initial excitation requirement. 
\end{remark}

\subsection{Stability Analysis}

\begin{theorem} \label{Thm1}
Provided that the auxiliary controller variable $v_r$ in \eqref{controlInput} is designed in the form
\begin{equation}
v_r = k_{n1} \rho_2^2 r + k_{n2}\left(  k_{\theta} N \lambda_d \overline{\rho}_1 \right) ^2 r \label{vr}
\end{equation}
with $k_{n1}$, $k_{n2} \in \mathbb{R}$ being positive damping gains selected to satisfy
\begin{eqnarray}
k_{n1} & \gg & \frac{1}{4 \min \left\lbrace \alpha \lambda_{\min} \left\lbrace K_p\right\rbrace , \lambda_{\min} \left\lbrace K_r\right\rbrace \right\rbrace} , \label{kn01} \\ 
k_{n2} & \gg & \frac{1}{4  k_{\theta}\underline{\lambda}} \label{kn02} 
\end{eqnarray}
the controller formulation proposed in \eqref{controlInput} with the gradient based plus concurrent learning type adaptation rule of \eqref{thetaCap} ensures global exponential tracking and exponential convergence of the estimation error in the sense that
\begin{equation}
\Vert y (t) \Vert \leq \left( \frac{\lambda_2}{\lambda_1}\right)^{\frac{1}{2}} \Vert y(0)\Vert \exp\left( -\frac{\gamma}{\lambda_2} t\right) \label{th01}
\end{equation}
where $y(t) \triangleq \left[ e^T \quad r^T \quad \tilde{\theta}^T \right]^T \in \mathbb{R}^{2n+p}$, $\gamma \in \mathbb{R}$ is a positive bounding constant and $\lambda_1$, $\lambda_2 \in \mathbb{R}$ are defined as
\begin{eqnarray}
\lambda_1 &\triangleq& \frac{1}{2} \min \Biggl\{ \lambda_{\min}\lbrace K_p \rbrace, m_1, \lambda_{\min}\lbrace \Gamma ^{-1} \rbrace \Biggl\}, \label{lamdasa} \\ 
\lambda_2 &\triangleq& \frac{1}{2} \max \Biggl\{ \lambda_{\max}\lbrace K_p \rbrace, m_2, \lambda_{\max}\lbrace \Gamma ^{-1} \rbrace \Biggl\}. \label{lamdasb}
\end{eqnarray}
\end{theorem}

\begin{proof} 
In order to prove Theorem \ref{Thm1}, we start by defining the following non negative function, $V(e,r,\tilde{\theta}): \mathbb{R}^n \times \mathbb{R}^n \times \mathbb{R}^p \rightarrow \mathbb{R}_+$ as
\begin{equation}
V \triangleq \frac{1}{2} e^T K_p e + \frac{1}{2} r^T M(q) r + \frac{1}{2}\tilde{\theta}^T \Gamma^{-1} \tilde{\theta}. \label{Vfsfb}
\end{equation}
Based on the structure of \eqref{Vfsfb} and Property \ref{prop1}, $V$ is bounded in the following form
\begin{equation}
\lambda_1 \Vert y \Vert^2 \leq V(y(t),t) \leq \lambda_2 \Vert y \Vert^2 \label{Vfsfbbound}
\end{equation} 
where $y(t)$ was previously introduced in \eqref{th01} and the bounding constants $\lambda_1$, $\lambda_2$ were defined in \eqref{lamdasa} and \eqref{lamdasb}, respectively.

Taking the time derivative of \eqref{Vfsfb} along \eqref{r}, \eqref{clR}, \eqref{thetaCl} yields
\begin{eqnarray}
\dot{V} &=& e^T K_p \left( -\alpha e + r \right) \nonumber \\ 
&& + \frac{1}{2} r^T \dot{M} r + r^T \left[ - C(q,\dot{q})r + Y_d \tilde{\theta} + \tilde{W} - K_r r -K_p e - v_r \right] \nonumber \\ 
&& + \tilde{\theta}^T \left( - Y_d^T r - k_{\theta} \sum_{i=1}^{N} \mathcal{Y}_i^T \Omega_i - k_{\theta} \sum_{i=1}^{N} \mathcal{Y}_i^T \mathcal{Y}_i \tilde{\theta} \right) . \label{V01_sil}
\end{eqnarray}
Applying \eqref{prp_2_1} of Property \ref{prop2} and canceling common terms results in
\begin{equation}
\dot{V} = - \alpha e^T K_p e + r^T \left( - K_r r + \tilde{W} - v_r \right) - \tilde{\theta}^T \left( k_{\theta} \sum_{i=1}^{N} \mathcal{Y}_i^T \Omega_i + k_{\theta} \sum_{i=1}^{N} \mathcal{Y}_i^T \mathcal{Y}_i \tilde{\theta} \right)
\end{equation}
which can be upper bounded using \eqref{WtildeBound}, \eqref{boundYdfS}, \eqref{underLamda} as
\begin{eqnarray}
\dot{V} &\leq& - \alpha \lambda_{\min} \left\lbrace K_p \right\rbrace \Vert e \Vert ^2 - \lambda_{\min} \left\lbrace K_r \right\rbrace \Vert r \Vert ^2 - k_{\theta} \underline{\lambda} \Vert \tilde{\theta} \Vert ^2 \nonumber \\ 
&& - \left[ k_{n1} \rho_2^2 \Vert r \Vert ^2 - \rho_2 \Vert r \Vert \, \Vert x \Vert \right] \nonumber \\
&& - \left[ k_{n2} \left( k_{\theta} \lambda_d N \overline{\rho}_1 \right)^2 \Vert r \Vert ^2 - k_{\theta} \lambda_d N \overline{\rho}_1 \Vert r \Vert \, \Vert \tilde{\theta} \Vert \right] \label{VdotBound01}
\end{eqnarray}
where \eqref{vr} was substituted as well. After completing the squares for the bracketed terms of \eqref{VdotBound01} and grouping common terms, we obtain 
\begin{equation}
\dot{V} \leq - \left[ \min \left\lbrace \alpha \lambda_{\min} \left\lbrace K_p \right\rbrace \, , \lambda_{\min} \left\lbrace K_r \right\rbrace \right\rbrace - \frac{1}{4 k_{n1}} \right] \Vert x\Vert ^2 - \left( k_{\theta} \underline{\lambda} - \frac{1}{4 k_{n2}} \right) \Vert \tilde{\theta} \Vert^2
\end{equation} 
and when the auxiliary control gains $k_{n1}$ and $k_{n2}$ are selected to satisfy the gain conditions of \eqref{kn01} and \eqref{kn02} respectively, we obtain
\begin{equation}
\dot{V} \leq - \gamma \Vert y \Vert^2  \label{VdotUpper}
\end{equation}
where $\gamma$ was introduced in \eqref{th01}. From the structure of $V(y,t)$ given in \eqref{Vfsfb} and the upper bound of its time derivative obtained in \eqref{VdotUpper}, we can invoke Theorem 4.10 of \cite{khalil} to prove that $e(t)$, $r(t)$ and $\tilde{\theta}(t)$ are exponentially convergent. Furthermore, substituting the second inequality of \eqref{Vfsfbbound} to the right hand side of \eqref{VdotUpper} and then solving the resulting differential inequality yields \eqref{th01}. Following standard signal chasing arguments, the boundedness of all closed loop signals are also guaranteed.
\end{proof}

%/************************************** \\
%yanliz proof semi-global... Burada Remark verip durumu aciklariz.. \\
%**************************************/
\begin{remark}
We would like to note that, as presented  in Appendix I, the bound of the term $\Omega(t)$ given in \eqref{OmegaBoundFsfb} contains quadratic terms. Therefore in order to damp these terms the corresponding controller gains would depend on the initial values of state dependent terms (\textit{i.e.} $e(0)$ and $r(0)$). This type of stability results are referred as \textit{semi}--global in control literature. For the details of semi--global type proofs, readers are referred to \cite{Lozano91} or \cite{compositeRob99}. 
\end{remark}

\section{Output Feedback Extension}

%/*****************************************\\
%Amanin ... bu algoritma desired version of state bagli ya biz bunu ofb de cozeriz... nasi mi ? aha böle... hemde hem gradient base hem de composite adaptive olarak... deyip gostericez.. yanlız composite te adaptasyon kazanci sabit olunca exponansiye oluyor.... cok hızlı ogrenme olunca bu tip algoritmalar sorun cikartiyor diyor Slotine.. forgetting olayi gerek falan diyor.. o islere girip girmemekte suphelerin var... \\
%ama çinliye nasıl yapılır görsün demek için yapıcam .. çinli sabit kazanlı yapmış ... bu şekilde tam least squares in faydalarından yararlanamaz ama bizim kinde olur argumanı iyi olur.. proof bir miktar dikkat istiyor.. V nin üst sınır olayı tehlikeli!!!! \\ 
%*****************************************/ \\

The control design problem in this section is restricted by the unavailability of joint velocity measurements. 

In order to remove the velocity measurement dependency of the controller in Section \ref{fsfb}, similar to \cite{compositeRob99} and \cite{BurgRob97}, we start by defining a surrogate signal designed to capture the behavior of $\dot{e}(t)$, represented with $e_f(t) \in \mathbb{R}^n$, as
\begin{equation}
e_f = \omega - k e \label{ef}
\end{equation} 
where $\omega(t) \in \mathbb{R}^n$  is an auxiliary variable updated according to
\begin{equation}
\dot{\omega} = - (\alpha_3 + k \alpha_2) (\omega - ke) - (k \alpha_1 - \alpha_2) e \quad \quad \omega (0) = k e(0) \label{w}
\end{equation}
where $k$, $\alpha_1$, $\alpha_2$, $\alpha_3 \in \mathbb{R}$ are positive constant gains. Taking the time derivative of \eqref{ef} and substituting for \eqref{w}, we obtain the following dynamical relationship
\begin{equation}
\dot{e}_f = -\alpha_3 e_f + \alpha_2 e - k \eta \quad \quad e_f(0) = 0_{n\times 1} \label{efdot}
\end{equation}
where the variable $\eta\left(t\right) \in \mathbb{R}^n$ is defined as
\begin{equation}
\eta \triangleq \dot{e} + \alpha_1 e +\alpha_2 e_f.  \label{eta}
\end{equation}
The subsequent composite adaptive version of the output feedback controller formulation will also make use of the prediction error term, $\varepsilon(t)$, defined in \eqref{epsi01}. The auxiliary term $\Omega(t)$ defined in \eqref{Omega} can be upper bounded with respect to the newly introduced error terms along \eqref{e}, \eqref{ef} and \eqref{eta} to yield (see Appendix \ref{appremark} for details)
\begin{equation}
\Vert \Omega \Vert \leq \rho_3(\Vert z \Vert) \Vert z \Vert \label{OmegaOfb}
\end{equation}
where $\rho_3(\cdot)\in\mathbb{R}$ is a positive definite bounding function and $z(t) \triangleq \left[ e^T \, e_f^T \, \eta^T \right] ^T\in \mathbb{R}^{3n}$.

Taking the time derivative of \eqref{eta}, pre--multiplying the resulting expression by $M(q)$, then substituting \eqref{modal}, \eqref{efdot}, \eqref{eta}, and adding / subtracting the $C(q,\dot{q})\eta$ and $Y_d \theta$ terms, we obtain
\begin{equation}
M(q)\dot{\eta} = -C(q,\dot{q}) \eta - \alpha_2 k M(q) \eta + Y_d \theta + \chi -\tau \label{openloopOFB01}
\end{equation}
where the auxiliary function $\chi(e, e_f, \eta) \in \mathbb{R}^n$ is defined as
\begin{eqnarray} 
\chi &\triangleq& M(q)\ddot{q}_d + F_d \dot{q} + G(q) - Y_d\theta \nonumber \\ 
     && + \alpha_1 M(q) (\eta -\alpha_1 e - \alpha_2 e_f) \nonumber \\ 
     && + \alpha_2 M(q) (-\alpha_3 e_f +\alpha_2 e) \nonumber \\ 
     && - C(q,\eta) (\dot{q}_d+\alpha_1 e + \alpha_2 e_f)  \nonumber \\ 
     && + C(q, \dot{q}_d)( \dot{q}_d + \alpha_1 e + \alpha_2 e_f) \nonumber  \\ 
     && + C(q, \dot{q}_d +\alpha_1 e + \alpha_2 e_f) (\alpha_1 e + \alpha_2 e_f) . \label{chiBB} 
\end{eqnarray}
Via utilizing Properties \ref{prop3}, \ref{DynamicBounds} and \eqref{eta}, the expression given in \eqref{chiBB} can be upper bounded as (details can be reached from in Appendix B of \cite{marcio-book})
\begin{equation}
\Vert \chi \Vert \leq \rho_4 (\Vert z \Vert) \Vert z \Vert \label{chiBound}
\end{equation}
where $\rho_4(\cdot)\in\mathbb{R}$ is a positive definite bounding function. Based on the obtained open loop error dynamics and the subsequent analysis, the control input torque is designed as
\begin{equation}
\tau = Y_d \hat{\theta} - k K_s e_f + K_s e \label{tauOFB}
\end{equation}
where $K_s \in \mathbb{R}^{n \times n}$ is a diagonal, positive definite, constant control gain matrix, and $Y_d$ was previously defined in Property \ref{P5}.

At this state, we will propose two different parameter update rules for the parameter estimation vector $\hat{\theta}(t)$. The first one is a standard gradient based update rule plus concurrent learning term and second is a composite adaptation rule formed via combining a gradient based estimator driven by the joint position tracking error and a least--squares term driven by prediction error term plus the concurrent learning term. 

The standard gradient based plus concurrent learning term type parameter estimation rule, in its joint velocity independent (\textit{i.e.}, implementable) form, is designed as 
\begin{eqnarray}
\hat{\theta} &=& \Gamma Y_d^T e + \Gamma \int_0^t Y_d^T(\sigma) \left[ \alpha_1 e (\sigma) + \alpha_2 e_f(\sigma) \right] d\sigma \nonumber \\ 
              && -\Gamma  \int_0^t \frac{d}{d\sigma} \left( Y_d^T(\sigma) \right) e(\sigma) d\sigma \nonumber \\ 
              && +  k_{\theta} \Gamma \int_0^t \sum_{i=1}^{N} \mathcal{Y}_i^T \biggl( \mathcal{U}_i - \mathcal{Y}_i \hat{\theta}  \biggr) d\sigma \label{thetaGradOfb}     
\end{eqnarray}
where $\mathcal{Y}_i$ and $\mathcal{U}_i$ were previously defined in \eqref{aha01} and \eqref{aha02}, respectively. Note that the time derivative of \eqref{thetaGradOfb} would yield
\begin{equation}
\dot{\hat{\theta}} = \Gamma Y_d^T \eta +  k_{\theta} \Gamma \sum_{i=1}^{N} \mathcal{Y}_i^T \biggl( \mathcal{U}_i - \mathcal{Y}_i \hat{\theta}  \biggr) \label{dotThetaGrad}
\end{equation}
where \eqref{eta} was used. 

For the composite adaptation plus concurrent learning based parameter estimation, the joint velocity independent form of the update rule is in the following structure  
\begin{eqnarray}
\hat{\theta} &=& P Y_d^T e + \int_0^t \left[ P(\sigma) Y_{df}^T(\sigma)\varepsilon(\sigma) + P(\sigma) Y_d^T(\sigma) \left( \alpha_1 e(\sigma) + \alpha_2 e_f(\sigma)\right) \right] d\sigma \nonumber \\ 
&& -\int_0^t \left[ -P(\sigma) Y_{df}^T(\sigma) Y_{df}(\sigma) P(\sigma) Y_{d}^T(\sigma) e(\sigma) + P(\sigma) \frac{d}{d\sigma} \left( Y_d^T(\sigma) \right) e(\sigma) \right] d\sigma \nonumber \\ 
&& -\int_0^t \left[ -P(\sigma) Y_{df}^T(\sigma) Y_{df}(\sigma) P(\sigma) Y_{d}^T(\sigma) e(\sigma) + P(\sigma)\dot{Y}_d^T(\sigma) e(\sigma) \right] d\sigma \nonumber \\ 
&& + k_{\theta} \int_0^t P(\sigma) \sum_{i=1}^{N} \mathcal{Y}_i^T \biggl( \mathcal{U}_i - \mathcal{Y}_i \hat{\theta} \biggr) d\sigma \label{thetaCompOfb}
\end{eqnarray}
where the prediction error term $\varepsilon(t)$ was previously defined in \eqref{epsi01} and $P(t) \in \mathbb{R}^{p \times p}$ is a time--varying gain matrix updated according to 
\begin{equation}
\dot{P} = -P Y_{df}^T Y_{df} P \label{P}
\end{equation}
so when $P(0) $ is selected to be a positive define and symmetric, $P(t)$ will remain positive definite and symmetric at all times. From \eqref{P} and $\frac{d}{dt} \left( P^{-1} \right) = - P^{-1} \dot{P} P^{-1}$, the time derivative of the inverse of $P$ is obtained as $\frac{d}{dt} \left( P^{-1} \right) = Y_{df}^T Y_{df}$ which will be positive definite and symmetric provided that $P(0)$ is selected positive definite and symmetric. Similar to its gradient based counterpart, the time derivative of \eqref{thetaCompOfb} is obtained as
\begin{equation}
\dot{\hat{\theta}} = P Y_{df}^T \varepsilon + P Y_d^T \eta + k_{\theta} P \sum_{i=1}^{N} \mathcal{Y}_i^T \biggl( \mathcal{U}_i - \mathcal{Y}_i \hat{\theta} \biggr). \label{dotThetaComp}
\end{equation}

The closed loop dynamics for $\eta$ can now be obtained by substituting the control torque input designed in \eqref{tauOFB} into the open loop dynamics of \eqref{openloopOFB01} to reach
\begin{equation}
M(q) \dot{\eta} = -C(q, \dot{q}) \eta - \alpha_2 k M(q) \eta + Y_d \tilde{\theta} + \chi + k K_s e_f -K_s e . \label{etaCL}
\end{equation} 
For the stability and convergence analysis of the error signals, we will present two different proofs with slightly different gain conditions for the concurrent learning adaptations with gradient based in \eqref{thetaGradOfb} and composite type in \eqref{thetaCompOfb} update rules.

\subsection{Gradient Adaptation Based Concurrent Learning}

\begin{theorem}
When the controller gains $k$ and $K_s$ are designed in the following form
\begin{eqnarray}
k &=& \frac{1}{m_1 \alpha_2} \left( \alpha_2 + k_{n2}  \left(  k_{\theta}\overline{\rho}_3 N \lambda_d \right)^2  + k_{n3} \rho_4^2 \right) \label{gradGains0} \\ 
\lambda_{\min} \left\lbrace K_s\right\rbrace  &\geq& 1 + k_{n2} \left(  k_{\theta} \overline{\rho}_3 N \lambda_d \right)^2 \label{gradGains}
\end{eqnarray}
where $\overline{\rho}_3 \in \mathbb{R}$ is defined as
\begin{equation}
\overline{\rho}_3 \triangleq \sup_{i=1,\cdots,N} \lbrace \rho_3 \left( \Vert z\left(t_i\right) \Vert \right) \rbrace ,
\end{equation}
the damping gain $k_{n2}$ satisfies \eqref{kn02} and $k_{n3}$ is selected to satisfy
\begin{equation}
k_{n3} \gg \frac{1}{4 \min \left\lbrace \alpha_1, \, \alpha_2, \, \alpha_3 \right\rbrace} \label{kn03}  
\end{equation}
then the velocity surrogate filter defined in \eqref{ef} and \eqref{w} and the controller proposed in \eqref{tauOFB} with the parameter estimation rule of \eqref{thetaGradOfb} ensure exponential convergence of the tracking error and on--line parameter estimation in the sense that
\begin{equation}
\Vert v(t) \Vert \leq \left( \frac{\lambda_4}{\lambda_3}\right)^{\frac{1}{2}} \Vert v(0) \Vert \exp\left( - \frac{\gamma_1}{\lambda_4}t \right)\quad \quad t \geq 0 \label{th02}
\end{equation}
where $v(t) = \left[e^T \, e_f^T \, \eta^T \, \tilde{\theta}^T \right]^T \in \mathbb{R}^{3n +p}$, $\gamma_1$ is a positive constant and $\lambda_3$, $\lambda_4 \in \mathbb{R}$ are defined as 
\begin{eqnarray}
\lambda_3 &\triangleq& \frac{1}{2} \min \left\lbrace \lambda_{\min} \left\lbrace K_s \right\rbrace, \, m_1 , \lambda_{\min}\left\lbrace \Gamma^{-1}\right\rbrace \right\rbrace  \label{gammasOfb0} \\ 
\lambda_4 &\triangleq& \frac{1}{2} \max \left\lbrace \lambda_{\max}\left\lbrace K_s \right\rbrace, \, m_2 , \lambda_{\max}\left\lbrace \Gamma^{-1}\right\rbrace \right\rbrace \label{gammasOfb}
\end{eqnarray}
\end{theorem}

\begin{proof}
We start our proof by defining a positive definite function $V_1(e,e_f,\eta, \tilde{\theta}): \mathbb{R}^n \times \mathbb{R}^n \times \mathbb{R}^n \times \mathbb{R}^p \rightarrow \mathbb{R}_+$ as
\begin{equation}
V_1 \triangleq \frac{1}{2}e^T K_s e + \frac{1}{2}e_f^T K_s e_f + \frac{1}{2} \eta^T M(q) \eta + \frac{1}{2} \tilde{\theta}^T \Gamma^{-1} \tilde{\theta} . \label{VOfb01}
\end{equation}
Note that, \eqref{VOfb01} can be bounded as follows
\begin{equation}
\lambda_3 \Vert v \Vert^2 \leq V_1(v(t),t) \leq \lambda_4 \Vert v \Vert^2
\end{equation}
where the bounding constants $\lambda_3$, $\lambda_4$ were defined in \eqref{gammasOfb0} and \eqref{gammasOfb}, respectively. 

Taking the time derivative of \eqref{VOfb01}, we obtain
\begin{eqnarray}
\dot{V}_1 &=& e^T K_s \left(-\alpha_2 e -\alpha_2 e_f + \eta \right) \nonumber \\ 
&& + e_f^T K_s  \left(-\alpha_3 e_f +\alpha_2 e - k \eta \right) \nonumber \\ 
&& + \frac{1}{2}\eta^T \dot{M}(q) \eta + \eta^T \left( -C(q,\dot{q}) \eta - \alpha_2 k M(q) \eta + Y_d \tilde{\theta} + \chi + k K_s e_f - K_s e \right) \nonumber \\
&& - \tilde{\theta}^T \left( Y_d^T \eta + k_{\theta} \sum_{i=1}^{N} \mathcal{Y}_i^T \biggl( \mathcal{U}_i - \mathcal{Y}_i \hat{\theta} \biggr) \right) . \label{VdotOfb01_sil} 
\end{eqnarray}
Canceling common terms, applying \eqref{epsi02} and Property \ref{prop2}, we obtain
\begin{eqnarray}
\dot{V}_1 &=& -\alpha_1 e^T K_s e - \alpha_3 e_f^T K_s e_f - \alpha_2 k \eta^T M(q)\eta + \eta^T \chi \nonumber \\ 
&& - \tilde{\theta}^T \left( k_{\theta} \sum_{i=1}^{N} \mathcal{Y}_i^T \Omega_i + k_{\theta} \sum_{i=1}^{N} \mathcal{Y}_i^T \mathcal{Y}_i \tilde{\theta} \right) \label{VdotOfb02}
\end{eqnarray}
which can be upper bounded in the form 
\begin{eqnarray}
\dot{V}_1 &\leq& -\alpha_1 \lambda_{\min}\lbrace K_s\rbrace \Vert e \Vert^2 - \alpha_3  \lambda_{\min}\lbrace K_s\rbrace \Vert e_f \Vert^2 - \alpha_2 k m_1 \Vert \eta \Vert^2 - k_{\theta} \underline{\lambda} \Vert\tilde{\theta}\Vert^2 \nonumber \\ 
&& + \eta^T \chi + k_{\theta} N \lambda_d \overline{\rho}_3 \Vert \tilde{\theta} \Vert \Vert z \Vert . \label{VdotOfb02Bound}
\end{eqnarray}
In order to further upper bound the right hand side of \eqref{VdotOfb02Bound}, we apply Property \ref{prop1}, \eqref{OmegaOfb}, \eqref{chiBound}, and group similar terms to obtain
\begin{eqnarray}
\dot{V}_1 &\leq& -\alpha_1 \Vert e \Vert^2 - \alpha_2 \Vert \eta \Vert^2 - \alpha_3 \Vert e_f \Vert^2 - k_{\theta} \underline{\lambda} \Vert\tilde{\theta}\Vert^2  \nonumber \\ 
           && -  \left[ k_{n2} \left(k_{\theta} N \lambda_d \overline{\rho}_3 \right) ^2  \Vert z \Vert ^2 - k_{\theta} N \lambda_d \overline{\rho}_3 \Vert z \Vert \, \Vert \tilde{\theta} \Vert \right] \nonumber \\ 
           && - \left[ k_{n3} \rho_4^2 \Vert\eta\Vert^2 - \rho_4\Vert\eta\Vert \Vert z \Vert \right] \label{VdotOfb03Bound}
\end{eqnarray}
where \eqref{gradGains0} and \eqref{gradGains} were substituted as well. After completing the squares of the bracketed terms of \eqref{VdotOfb03Bound}, we obtain
\begin{equation}
\dot{V}_1 \leq - \left( \min \left\lbrace \alpha_1, \, \alpha_2, \, \alpha_3 \right\rbrace - \frac{1}{4k_{n3}} \right) \Vert z \Vert ^2 -  \left( k_{\theta} \underline{\lambda} - \frac{1}{4 k_{n2}} \right) \Vert \tilde{\theta} \Vert ^2
\end{equation} 
and when the damping gains $k_{n2}, k_{n3}$ are selected to satisfy \eqref{kn02} and \eqref{kn03}, respectively, we obtain
\begin{equation}
\dot{V}_1 \leq - \gamma_1 \Vert v \Vert^2
\end{equation}
where $\gamma_1$ was introduced in \eqref{th02}. Following the footsteps of Theorem \ref{Thm1}, we invoke Theorem 4.10 of \cite{khalil} to prove \eqref{th02} and thus we conclude that $e(t)$ and $\tilde{\theta}(t)$ are exponentially convergent to the origin, and all the signals under the closed loop system remain bounded.
\end{proof}

\subsection{Composite Adaptation Based Concurrent Learning}

\begin{theorem}\label{thm3}
The controller of \eqref{tauOFB} in conjunction with the parameter update rule of \eqref{thetaCompOfb} and the least--squares gain matrix update law in \eqref{P} with the velocity surrogate filter defined in \eqref{ef} and \eqref{w} ensures global uniform asymptotic stability and convergence of the signals $e(t)$ and $\tilde{\theta}(t)$ to origin provided that the controller gains $k$ and $K_s$ of \eqref{tauOFB} are selected to satisfy
\begin{eqnarray}
k &=& \frac{1}{m_1 \alpha_2} \left( \alpha_2 + k_{n2}  \left( k_{\theta} N \lambda_d \overline{\rho}_3 \right)^2 + k_{n3} \rho_4^2 + k_{n4} \rho_3^2 \right) \label{compGains0} \\ 
\lambda_{\min} \left\lbrace K_s\right\rbrace  &\geq& 1 + k_{n2} \left( k_{\theta} N \lambda_d \overline{\rho}_3 \right)^2 + k_{n4} \rho_3^2 \label{compGains}
\end{eqnarray}
where $k_{n2}$, $k_{n3}$ satisfy \eqref{kn02} and \eqref{kn03}, respectively, and the newly introduced damping gain $k_{n4} \in \mathbb{R}$ is selected to satisfy
\begin{equation}
k_{n4} \gg \frac{1}{2} . \label{kn04}
\end{equation}
\end{theorem}

\begin{proof}
In order to prove the theorem, we start by defining $V_2(e,e_f,\eta, \tilde{\theta}) : \mathbb{R}^n \times \mathbb{R}^n \times \mathbb{R}^n \times \mathbb{R}^p \rightarrow \mathbb{R}_+$ as follows
\begin{equation}
V_2 \triangleq \frac{1}{2}e^T K_s e + \frac{1}{2}e_f^T K_s e_f + \frac{1}{2} \eta^T M(q) \eta + \frac{1}{2} \tilde{\theta}^T P^{-1} \tilde{\theta} \label{VOfb02}
\end{equation} 
where $P$ is the time--varying adaptation gain matrix updated according to \eqref{P}. Note that $V_2(v,t)$ of \eqref{VOfb02}, is positive definite, and radially unbounded. 

After taking the time derivative of \eqref{VOfb02}, inserting for \eqref{epsi02}, \eqref{efdot}, \eqref{eta}, \eqref{P}, \eqref{dotThetaComp}, \eqref{etaCL}, canceling common terms and applying Property \ref{prop2}, we obtain
\begin{eqnarray}
\dot{V}_2 &=& -\alpha_1 e^T K_s e - \alpha_3 e_f^T K_s e_f - \alpha_2 k \eta^T M(q)\eta + \eta^T \chi \nonumber  \\ 
    && -\frac{1}{2}\Vert Y_{df}\tilde{\theta} \Vert^2 -\tilde{\theta}^T Y_{df}^T \Omega -\tilde{\theta}^T \left( k_{\theta} \sum_{i=1}^{N} \mathcal{Y}_i^T \Omega_i + k_{\theta} \sum_{i=1}^{N} \mathcal{Y}_i^T \mathcal{Y}_i \tilde{\theta} \right) \label{VdotOfbComp}
\end{eqnarray}
which can be upper bounded in the form 
\begin{eqnarray}
\dot{V}_2 &\leq& -\alpha_1 \lambda_{\min} \lbrace K_s\rbrace \Vert e \Vert^2 - \alpha_3  \lambda_{\min}\lbrace K_s\rbrace \Vert e_f \Vert^2 - \alpha_2 k m_1 \Vert \eta \Vert^2 - k_{\theta} \underline{\lambda} \Vert\tilde{\theta}\Vert^2 \nonumber \\ 
            && -\frac{1}{2}\Vert Y_{df}\tilde{\theta} \Vert^2  +\Vert Y_{df}\tilde{\theta} \Vert\Vert \Omega \Vert +  \eta^T \chi + k_{\theta} N \lambda_d \overline{\rho}_3 \Vert \tilde{\theta} \Vert \Vert z \Vert . \label{VdotOfbCompBound}
\end{eqnarray}
Applying Property \ref{prop1}, and making use of the gain definitions of \eqref{compGains0} and \eqref{compGains}, we further bound \eqref{VdotOfbCompBound} as
\begin{eqnarray}
\dot{V}_2 &\leq& -\alpha_1 \Vert e \Vert^2 - \alpha_2 \Vert \eta \Vert^2 - \alpha_3 \Vert e_f \Vert^2 - k_{\theta} \underline{\lambda} \Vert\tilde{\theta}\Vert^2  \nonumber \\ 
           && - \left[ k_{n2} \left( k_{\theta} N \lambda_d \overline{\rho}_3 \right) ^2  \Vert z \Vert ^2 - k_{\theta} N \lambda_d \overline{\rho}_3 \Vert z \Vert \, \Vert \tilde{\theta} \Vert \right] \nonumber \\ 
           && - \left[ k_{n3} \rho_4^2 \Vert\eta\Vert^2 - \rho_4\Vert\eta\Vert \Vert z \Vert \right] \nonumber  \\
           && - \frac{1}{2}\Vert Y_{df}^T\tilde{\theta} \Vert^2 - \left[ k_{n4} \rho_3^2 \Vert z \Vert ^2 - \rho_3 \Vert z \Vert \Vert Y_{df}^T\tilde{\theta} \Vert \right]  . \label{VdotOfbCompBound02}
\end{eqnarray}
After completing the squares of the bracketed terms of \eqref{VdotOfbCompBound02}, we obtain
\begin{eqnarray}
\dot{V}_2 &\leq& - \left( \min \left\lbrace \alpha_1, \, \alpha_2, \, \alpha_3 \right\rbrace - \frac{1}{4k_{n3}} \right) \Vert z \Vert^2\nonumber \\
        & &  - \left( k_{\theta} \underline{\lambda} - \frac{1}{4 k_{n2}} \right) \Vert \tilde{\theta} \Vert ^2 \nonumber \\
        & &  -  \left( \frac{1}{2} - \frac{1}{4 k_{n4}} \right) \Vert Y_{df}^T\tilde{\theta} \Vert^2
\end{eqnarray}
which, via utilizing \eqref{kn04}, can further be upper bounded to have the following form
\begin{equation}
\dot{V}_2 \leq - \left( \min \left\lbrace \alpha_1, \, \alpha_2, \, \alpha_3 \right\rbrace - \frac{1}{4k_{n3}} \right) \Vert z \Vert^2 
- \left( k_{\theta} \underline{\lambda} - \frac{1}{4 k_{n2}} \right) \Vert \tilde{\theta} \Vert ^2
\end{equation} 
and when the damping gains $k_{n2}$, $k_{n3}$ are selected to satisfy \eqref{kn02} and \eqref{kn03}, respectively, we obtain
\begin{equation}
\dot{V}_2 \leq - \gamma_2 \Vert v \Vert^2 \label{dotV2}
\end{equation}
for some positive constant $\gamma_2 \in \mathbb{R}$. As can be observed from the structures of \eqref{VOfb02} and \eqref{dotV2}, $V_2(v,t)$ is positive definite, decrescent and radially unbound with its time derivative, $\dot{V}_2(v,t)$ being negative definite, therefore uniform asymptotic convergence of $v(t)$ is established \cite{khalil}. As $P^{-1}(t)$ is positive definite for $t \geq 0$ we conclude that $P(t)$ is positive definite for all time. And since $P(t)$ is positive definite with  $\dot{P} \leq 0$ we have $P(t)$ bounded for all time. Standard signal chasing arguments can now be used to show all signals remain bounded in the closed loop system. 
\end{proof}
\begin{remark}
Note that when instead of the time varying adaptation gain matrix, $P(t)$ introduced in \eqref{thetaCompOfb}, a fixed valued adaptation matrix (similar to \cite{PanYu18}), like $\Gamma$ of \eqref{thetaGradOfb} were used, it is trivial that the stability result of Theorem \ref{thm3} would be exponential convergence. However the use of fixed valued adaptation gain in a composite estimator would not allow the full use of least squares based estimation term. Therefore we preferred a time varying adaptation gain in our study.   
\end{remark}
\begin{remark}
The composite adaption based concurrent learning extension presented for the output feedback controller can also be applied to the full state controller presented in Section \ref{fsfb}, however is omitted since it is trivial.  
\end{remark}

%\section{Numerical Studies}\label{simu}

%In order to demonstrate the performance of the proposed controller, \\
%/************************** \\
%Serhat in ellerinden öper.. \\
%Arxiv daki versiyonuna simulasyon olmasa da olur... \\
%**************************/ 

\section{Conclusions}\label{conclusion}

In this study, a novel approach is proposed as a solution to the problem of tracking control of Euler Lagrange systems while simultaneously identifying dynamic model parameters. Concurrent learning type adaptation is preferred to be used where the regression matrix utilized in the design of the adaptive update law is formed to rely on desired states which is a novel departure from the relevant results in the literature. Exponential tracking and parameter identification is ensured via Lyapunov methods. 

The use of desired states in the design of the adaptive update rule allowed the proposed control strategy to be extended for the output feedback problem as well. That is, via the design of a surrogate signal, the dependence of the control input torque to velocity measurements is eliminated. Two different adaptive update rules are proposed with the first one being gradient based (\textit{i.e.}, similar to the full state feedback case) and the second one being least squares type with time varying gain matrix. For both output feedback controllers, convergence of the tracking error and parameter identification error to the origin are ensured via novel Lyapunov tools.

\appendices

\section{Proof of Bounds} \label{appremark}

\begin{lemma}
The auxiliary error--like variable $\Omega(t)$ defined in \eqref{Omega} can be upper bounded with respect to $e(t)$, $r(t)$ (or $x(t)$) as given in \eqref{OmegaBoundFsfb} and with respect to $e(t)$, $e_f(t)$, $\eta(t)$ (or $z(t)$) as given in \eqref{OmegaOfb}.
\end{lemma}
\begin{proof}
In order to prove the above Lemma, we will make use of the standard properties of the convolution operation and the following bounding statement \cite{vidyasagar-book}
\begin{equation}
\Vert g_1 (t) * g_2(t) \Vert \leq \Vert g_1(t) \Vert \Vert g_2(t) \Vert \quad \forall g_1(t), g_2(t) \in \mathbb{R}^n. \label{convProp}
\end{equation} 
It is also noted that, following statements can be easily obtained for the filter function given in \eqref{f_filter} and its time derivative in \eqref{f_dot}
\begin{equation}
\vert f(0) \vert = \beta,  \quad \vert f(t) \vert \leq \beta, \quad \vert \dot{f}(t) \vert \leq \beta^2 . \label{UboundFs}
\end{equation}

To obtain the upper bound of $\Omega(t)$ in terms of $e(t)$ and $r(t)$, we apply \eqref{prp_2} of Property \ref{prop2} and reformulate some terms in \eqref{Omega2} to reach
\begin{eqnarray}
\Omega &=& \dot{f}(t) * \lbrace \left( M(q) - M(q_d) \right) \dot{q} - M(q_d) \dot{e} \rbrace \nonumber \\ 
&& + f(0) \left\lbrace \left( M(q) - M(q_d) \right) \dot{q} - M(q_d) \dot{e} \right\rbrace \nonumber \\ 
&& + f(t) * \left\lbrace - C^T(q, \dot{q}) \dot{q} + C^T(q_d, \dot{q}_d) \dot{q}_d + G(q) - G(q_d) + F_d \left( \dot{q} - \dot{q}_d \right) \right\rbrace \label{OmegaBound1}
\end{eqnarray}
where \eqref{prp_2} of Property \ref{prop2} and Assumption \ref{Ass1} were used as well. The first two terms in the bracket of the last line can be manipulated as follows
\begin{equation}
- C^T(q, \dot{q}) \dot{q} + C^T(q_d, \dot{q}_d) \dot{q}_d = \left[ C(q_d, \dot{q}_d) - C(q, \dot{q}_d) \right]^T \dot{q}_d + \left[ C(q, \dot{q}_d) - C(q, \dot{q}) \right]^T \dot{q}_d + C^T(q, \dot{q}) \left( \dot{q}_d - \dot{q} \right) \label{OmegaBound2}
\end{equation}
Substituting \eqref{OmegaBound2} into \eqref{OmegaBound1} and then utilizing \eqref{convProp} and \eqref{UboundFs} yields
\begin{eqnarray}
\Vert \Omega \Vert & \leq& \beta^2 \left( \Vert M(q) - M(q_d) \Vert_{i \infty} \Vert \dot{q} \Vert + \Vert M(q_d) \Vert_{i \infty} \Vert \dot{e} \Vert \right) \nonumber \\ 
&& + \beta \left( \Vert M(q) - M(q_d) \Vert_{i \infty} \Vert \dot{q} \Vert + \Vert M(q_d) \Vert_{i \infty} \Vert \dot{e} \Vert \right) \nonumber \\ 
&& + \beta \left( \Vert C(q_d, \dot{q}_d) - C(q, \dot{q}_d) \Vert_{i \infty} \Vert \dot{q}_d \Vert + \Vert C(q, \dot{q}_d) - C(q, \dot{q}) \Vert_{i \infty} \Vert \dot{q}_d \Vert \right. \nonumber \\
&& \left. + \Vert C(q, \dot{q}) \Vert_{i \infty} \Vert \dot{e} \Vert + \Vert G(q) - G(q_d) \Vert + \Vert F_d \Vert_{i \infty} \Vert \dot{e} \Vert \right) \label{OmegaBound3}
\end{eqnarray}
to which applying the bounds from Properties \ref{prop1}, \ref{prop3} and \ref{DynamicBounds} following can be obtained
\begin{eqnarray}
\Vert \Omega \Vert &\leq& \left( \beta^2 + \beta \right) \left( \zeta_{M1} \Vert e \Vert \Vert \dot{q} \Vert + m_2 \Vert \dot{e} \Vert \right) \nonumber \\
&& + \beta \left( \zeta_{C2} \Vert e \Vert \Vert \dot{q}_d \Vert^2 + \zeta_{C1} \Vert \dot{e} \Vert \Vert \dot{q}_d \Vert + \zeta_{C1} \Vert \dot{q} \Vert \Vert \dot{e} \Vert + \zeta_G \Vert e \Vert + \zeta_F \Vert \dot{e} \Vert \right) . \label{OmegaBound4}
\end{eqnarray}
In view of time derivative of \eqref{e}, $\Vert \dot{q} \Vert \leq \Vert \dot{q}_d \Vert + \Vert \dot{e} \Vert $ and since the desired trajectory and its time derivatives are bounded then $\Vert \dot{q}_d \Vert \leq \zeta_{d1}$ for some positive bound $\zeta_{d1} \in \mathbb{R}$. Utilizing these boundedness statements results in
\begin{eqnarray}
\Vert \Omega \Vert &\leq& \left( \beta^2 + \beta \right) \left( \zeta_{M1} \zeta_{d1} \Vert e \Vert + \zeta_{M1} \Vert e \Vert \Vert \dot{e} \Vert + m_2 \Vert \dot{e} \Vert \right) \nonumber \\
&& + \beta \left( \zeta_{C2} \zeta_{d1}^2 \Vert e \Vert + 2 \zeta_{C1} \zeta_{d1} \Vert \dot{e} \Vert 
+ \zeta_{C1} \Vert \dot{e} \Vert^2 + \zeta_G \Vert e \Vert + \zeta_F \Vert \dot{e} \Vert \right) . \label{OmegaBound4b}
\end{eqnarray}
From \eqref{r}, $ \Vert \dot{e} \Vert \leq \Vert r \Vert$ can be obtained \cite{lewis} and utilizing this upper bound along with \eqref{OmegaBound4b} yields
\begin{eqnarray}
\Vert \Omega \Vert &\leq& \left( \beta^2 + \beta \right) \left( \zeta_{M1} \zeta_{d1} \Vert e \Vert + \zeta_{M1} \Vert e \Vert \Vert r \Vert + m_2 \Vert r \Vert \right) \nonumber \\
&& + \beta \left( \zeta_{C2} \zeta_{d1}^2 \Vert e \Vert + 2 \zeta_{C1} \zeta_{d1} \Vert r \Vert + \zeta_{C1} \Vert r \Vert^2 + \zeta_G \Vert e \Vert + \zeta_F \Vert r \Vert \right) \label{OmegaBound5}
\end{eqnarray}
and utilizing the bound of $ \Vert e \Vert \leq \frac{1}{\alpha} \Vert r \Vert$ \cite{lewis} allows the right hand side to be upper bounded as
\begin{equation}
\Vert \Omega \Vert \leq \left( \xi_1 + \xi_2 \Vert e \Vert + \xi_3 \Vert r \Vert \right) \Vert r \Vert \label{OmegaBound6}
\end{equation}
for some positive bounds $\xi_1$, $\xi_2$, $\xi_3 \in \mathbb{R}$ and the bound of \eqref{OmegaBoundFsfb} can easily be reached via choosing $\rho_1 \triangleq \xi_1 + \max \lbrace \xi_2 , \xi_3 \rbrace \Vert x \Vert$.
% ***********************************************************************************************************
% BOUND UN DEVANI BURADA 
%\begin{eqnarray}
%\Vert \Omega \Vert &\leq& \left( \beta^2 + \beta \right) \left\lbrace 
%                      \zeta_{M1} \Vert \dot{q}_d \Vert \Vert e \Vert +\zeta_{M1} \Vert r \Vert \Vert e \Vert +\zeta_{M1} \alpha^2 \Vert e \Vert^2
%                      +  \zeta_{md} \Vert e \Vert \right\rbrace   \nonumber \\
%                   && + \beta \left\lbrace \zeta_{d0} + \zeta_{c1} \Vert \dot{q}_d \Vert ^2
%                      + \zeta_{c1} \Vert r \Vert ^2 + \zeta_{c1} \alpha^2 \Vert e \Vert ^2
%                   + \zeta_{cd} \Vert q_d\Vert + \zeta_G \Vert e \Vert + \zeta_F \Vert e \Vert \right\rbrace  \label{Oha01}
%\end{eqnarray}
%due to the quadratic nature with respect to $\Vert e \Vert $ and $ \Vert r \Vert$ of \eqref{Oha01} it can be upper bounded in the following form
%\begin{equation}
%\Vert \Omega \Vert \leq \rho_1 (e, r) \Vert r \Vert
%\end{equation} for some positive bounding function $\rho_1$
% ***********************************************************************************************************

When obtaining the upper bound of $\Omega(t)$ in terms of $e(t)$, $e_f(t)$ and $\eta(t)$, we continue from the bound in \eqref{OmegaBound4b}. From \eqref{eta}, $\Vert \dot{e} \Vert \leq \Vert \eta - \alpha_1 e -\alpha_2 e_f \Vert \leq \max \lbrace \alpha_1 , \alpha_2 \rbrace \Vert z \Vert$ can be obtained. Since $\Vert e \Vert \leq \Vert z \Vert$, then, from \eqref{OmegaBound4b}, it is straightforward to obtain the following expression
\begin{eqnarray}
\Vert \Omega \Vert &\leq& \left( \beta^2 + \beta \right) \left( \zeta_{M1} \zeta_{d1} \Vert z \Vert + \zeta_{M1} \max \lbrace \alpha_1 , \alpha_2 \rbrace \Vert z \Vert^2 + m_2 \max \lbrace \alpha_1 , \alpha_2 \rbrace \Vert z \Vert \right) \nonumber \\
&& + \beta \left( \zeta_{C2} \zeta_{d1}^2 \Vert z \Vert + 2 \zeta_{C1} \zeta_{d1} \max \lbrace \alpha_1 , \alpha_2 \rbrace \Vert z \Vert 
+ \zeta_{C1} \left[ \max \lbrace \alpha_1 , \alpha_2 \rbrace \right]^2 \Vert z \Vert^2 \right. \nonumber \\
&& \left. + \zeta_G \Vert z \Vert + \zeta_F \max \lbrace \alpha_1 , \alpha_2 \rbrace \Vert z \Vert \right) \label{OmegaBound7}
\end{eqnarray}
which, for some positive bounds $\xi_4$, $\xi_5 \in \mathbb{R}$, can be rewritten as
\begin{equation}
\Vert \Omega \Vert \leq \xi_4 \Vert z \Vert + \xi_5 \Vert z \Vert ^2 \label{geneOmega}
\end{equation}
where the bound of \eqref{OmegaOfb} can be obtained by choosing $\rho_3 \triangleq \xi_4 + \xi_5 \Vert z \Vert$.
\end{proof}

\bibliographystyle{IEEEtran}
\bibliography{IEEEabrv,Refs}

\end{document}